# The investigation of $YAlO_3$-$NdAlO_3$ system, synthesis and characterization

A. Szysiak[1], D. Klimm[2], S. Ganschow[2], M. Mirkowska[1], R. Diduszko[1], L. Lipińska[1], A. Kwasniewski[2], A. Pajączkowska[1]

[1] Institute of Electronic Materials Technology, Wólczyńska 133, 01-919 Warsaw, Poland
[2] Leibniz Institute for Crystal Growth, Max-Born Str. 2, 12489 Berlin, Germany

**Abstract**

The binary phase diagram of the $YAlO_3$ (YAP) – $NdAlO_3$ (NAP) system was determined by differential thermal analysis (DTA) and X-ray powder diffraction (XRD) measurements. High purity nanocrystalline powders and small single crystals of $Y_{1-x}Nd_xAlO_3$ ($0 \leq x \leq 1$) have been produced successfully by modified sol-gel (Pechini) and micro-pulling-down methods, respectively. Both end members show high mutual solubility > 25% in the solid phase, with a miscibility gap for intermediate compositions. A solid solution with $x \approx 0.2$ melts azeotropic ca. 20 degrees below pure YAP. Such crystals can be grown from the melt without segregation. The narrow solid/liquid region near the azeotrope point could be measured with a "cycling" DTA measurement technique.

## 1. Introduction

Yttrium aluminum perovskite (YAP) doped with rare-earth or transition metal ions is a very promising material because of its optical properties. It has been successfully used as a host material for solid-state lasers[1,2], scintillators[3,4], phosphors[5,6], high output power resonators[7] or even ceramic pigments[8]. On the other hand, neodymium aluminum perovskite (NAP) seems to be an interesting material which is used as a component of nanocomposites with alumina to improve luminescence efficiency itself[9]. As these two perovskites have different structures, YAP orthorhombic[10] and NAP rhombohedral[11], unlimited mutual solubility of both phases is not possible. To the author's knowledge the binary phase diagram YAP-NAP and the melting behavior of mixtures $Y_{1-x}Nd_xAlO_3$ ($0 \leq x \leq 1$) were not yet reported in the literature except of pure or weakly (3 mass%)[12] $Nd^{3+}$ doped YAP, and for pure NAP ($x = 1$). Pure NAP melts congruently near 2100°C[13]. Pure YAP is reported by several authors to melt congruently near 1900°C, but the phase was reported to be unstable for $T < 1600°C$[14]. It was reported, however, that YAP is stable if annealed as long as 300 hours in air at 1200°C, and melts congruently at 1940°C[15]. This observation is in agreement with a recent assessment of the $Al_2O_3$-$Y_2O_3$ system where YAP was found to be stable down to low temperatures[16]. YAP with limited amount of dopant can be grown as a single crystal by the Czochralski technique but obtaining it as crystalline powder by wet chemical method is very difficult[17,18].

The purpose of this work is the investigation of YAP-NAP binary phase diagram. The solubility of Nd in the $YAlO_3$ structure was checked by obtaining suitable materials by sol-gel and micro-pulling-down methods. The focus will be on high temperature phase equilibria near the solidus temperatures that are relevant for crystal growth from the melt.

## 2. Experimental

Nanocrystalline powders of $Y_{1-x}Nd_xAlO_3$ ($0 \leq x \leq 1$) were synthesized by modified sol-gel (Pechini) method described in detail in our previous publication[19]. The regime of calcination was following: when powders were calcined in air even at 1000°C for 12 h the single-phase neodymium aluminum perovskite was created. It was found, however, that this temperature was not sufficient for obtaining single-phase yttrium aluminum perovskite. After subsequent annealing at 1600°C for 7 h, single-phase $YAlO_3$ was obtained.

$Y_{1-x}Nd_xAlO_3$ single crystals have been grown using the micro-pulling-down technique with rf induction heating. Starting materials were (a) $YAlO_3$ and $NdAlO_3$ powder mixtures prepared from high purity oxides by solid state reaction, and (b) nanocrystalline powders prepared by sol-gel synthesis as described above. The material was melted in an iridium crucible which was thermally insulated by zirconia and alumina ceramics. The growth chamber was continuously rinsed with protective gas ($N_2$). Crystals were pulled with a rate of 0.20 mm/min starting on an iridium wire. In all experiments the entire material input (approximately 0.5 g) was crystallized.

The samples were investigated by a variety of analytical methods: Crystal lattice parameters of nanocrystalline powders as well as powdered single crystals were measured by X-ray diffraction (XRD) using Siemens D-500 diffractometer with $CuK_\alpha$ radiation at 1.548 Å. A Rietveld analysis of the X-ray spectra was performed with computer program PowderCell for Windows Version 2.4. The grain size and surface morphology were analyzed by high resolution scanning electron microscopy (HRSEM) with LEO GEMINI 1530 apparatus. Differential thermal analysis (DTA) measurements were performed with a NETZSCH STA 449 F1 "Jupiter" (graphite furnace, flowing argon atmosphere, up to 2000°C), or with a STA 429 CD (tungsten furnace, stationary helium atmosphere, up to 2400°C). For all measurements tungsten crucibles and tungsten sample holders (W/Re thermocouples) were used, and heating/cooling rates ±15 deg/min were applied.

## 3. Results and Discussion

Pure YAP crystallizes in the orthorhombic space group *Pbnm* (number 62 of the International Tables), and the X-ray pattern is given in ICDD 01-089-7947. The orthorhombic symmetry arises from a distortion of the ideal cubic perovskite archetype along a ⟨110⟩ direction, and results in a magnification of the unit cell volume by a factor of 4. In contrast, the rhombohedral unit cell of NAP with space group $R\bar{3}c$ (number 167, ICDD 04-007-8024) is the result of a distortion of the cubic unit cell along the space diagonal ⟨111⟩ and is 6 times larger.

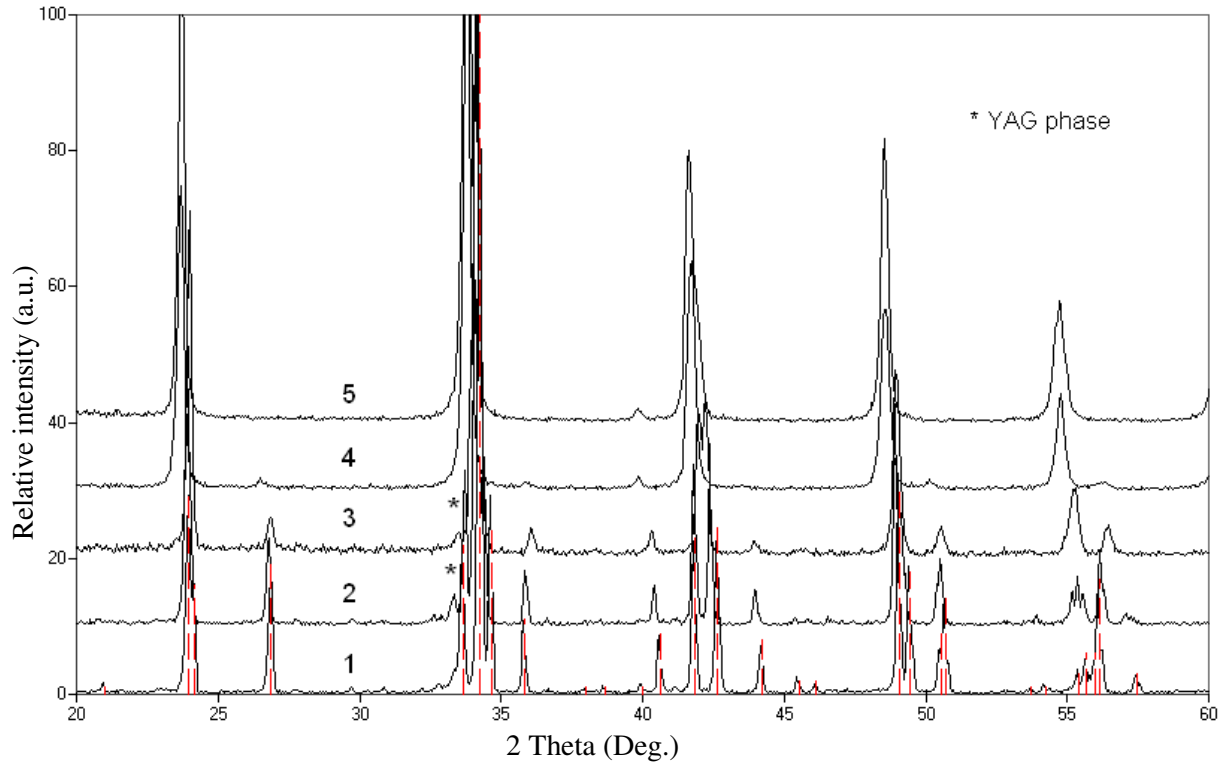

**Fig. 1: X-ray diffraction patterns for 5 annealed nanocrystalline powder samples ranging from YAP to NAP: 1) YAlO$_3$; 2) Y$_{0.8}$Nd$_{0.2}$AlO$_3$; 3) Y$_{0.55}$Nd$_{0.45}$AlO$_3$; 4) Y$_{0.2}$Nd$_{0.8}$AlO$_3$; 5) NdAlO$_3$.**

The crossover from the orthorhombic YAP to the rhombohedral NAP can be seen from the X-ray diffraction patterns for selected samples that are shown in Fig. 1. Only samples on the orthorhombic YAP side show the garnet (YAG) related diffraction peak near 33° which vanishes if this phase disappears. It means that for the YAlO$_3$ doped with neodymium ions, except main phase of perovskite the residual amount of garnet (Y$_3$Al$_5$O$_{12}$) is present. This result could be explained by the instability of YAP at low $T^{14}$.

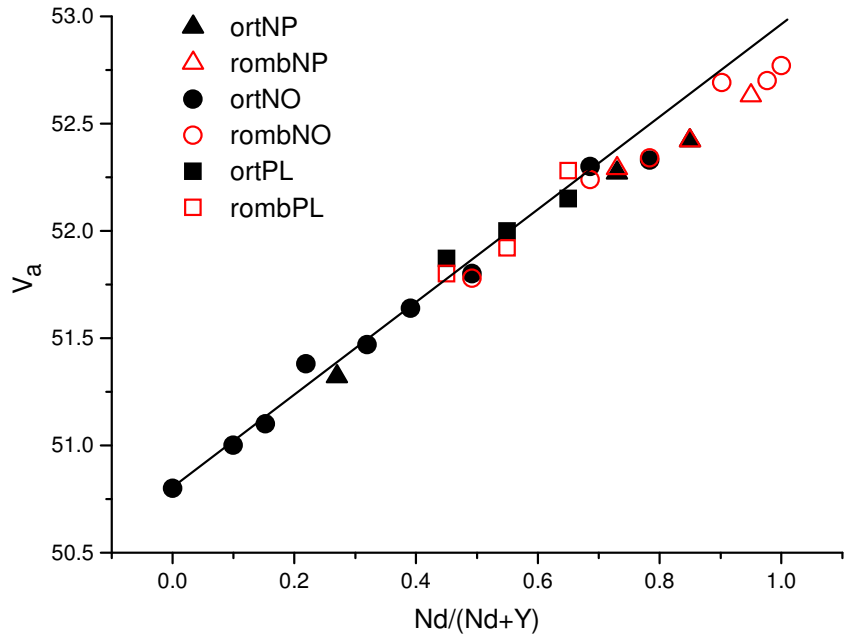

**Fig. 2: The volume $V_a$ of the perovskite archetype cell (1 formula unit) depending on molar ratio Nd/(Nd+Y). Filled symbols – orthorhombic structure, hollow symbols – rhombohedral structure; triangular symbols – samples obtained by micro-pulling-down method from nanopowders, circular symbols – samples obtained by micro-pulling-down method from oxides, square symbols – nanopowders obtained by modified sol-gel method.**

The results of detailed XRD investigations are presented in Fig. 2. The changes of the volume $V_a$ of the perovskite archetype cell depend on molar ratio Nd/(Nd+Y) and indicate that from about 0.45 to 0.80 both orthorhombic and rhombohedral phases are present. For the comparison of changes of geometric parameters versus ionic contribution for every structure the volume of unit cell was converted to volume $V_a$ of the archetype (quasi-cubic) perovskite unit cell that contains one $AlO_6$ octahedron only. From Fig. 2 it is obvious that $V_a$ increases with the content of $Nd^{3+}$ ions because the $Nd^{3+}$ ion (1.12 Å) is larger than $Y^{3+}$ (1.04 Å). Particularly at lower content of neodymium $x$ = Nd/(Nd+Y), $V_a$ depends linearly from the concentration. The linearity is observed for $0 \leq x \leq 0.7$. This limit concentration corresponds to an average ionic radius of $0.3 \times 1.04$ Å + $0.7 \times 1.12$ Å = 1.096 Å, which is almost identical with the octahedral Shannon radius of samarium (1.098 Å)[20]. It should be remarked that all $REAlO_3$ with smaller rare earth ions up to $SmAlO_3$ are reported to be orthorhombic, and for the larger RE from Nd on these perovskites are rhombohedral[11]. Obviously the crossover from orthorhombic to rhombohedral distortion of the cubic archetype unit cell depends only from the average ionic radius of the RE. The calculated volume of archetypical cells for the same nominal composition for the both structures is very similar.

**Table 1: The lattice parameters of 4 annealed powder samples ranging from YAP to NAP.**

| Crystal | $YAlO_3$ | $Y_{0.45}Nd_{0.55}AlO_3$ | $Y_{0.05}Nd_{0.95}AlO_3$ | $NdAlO_3$ |
|---|---|---|---|---|
| Crystal structure | $Pnma$ orthorhombic | $Pnma$ orthorhombic | $R\bar{3}c$ rhombohedral | $R\bar{3}c$ rhombohedral |
| Unit cell dimensions (Å) | a = 5.180<br>b = 5.320<br>c = 7.375 | a = 5.269<br>b = 5.302<br>c = 7.465 | a = 5.321<br>c = 12.915 | a = 5.321<br>c = 12.916 |

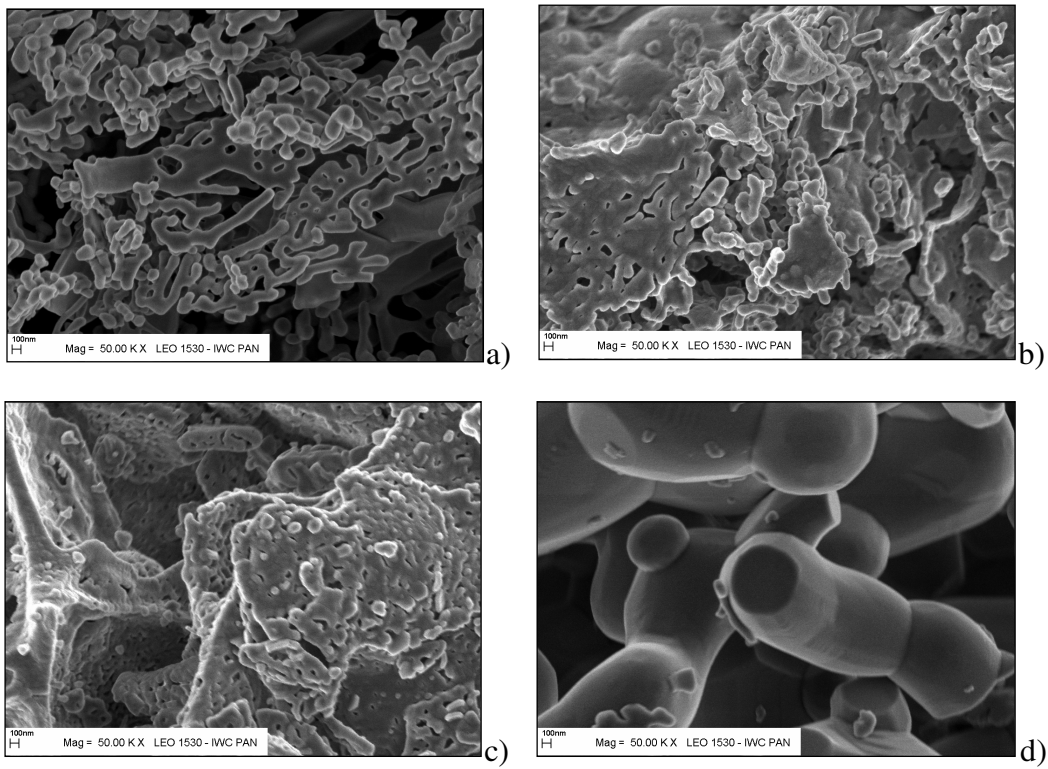

**Fig. 3: HRSEM images of morphology of nanopowders. The Nd concentration decreases from top left to bottom right: a) $Y_{0.05}Nd_{0.95}AlO_3$; b) $Y_{0.15}Nd_{0.85}AlO_3$; c) $Y_{0.45}Nd_{0.55}AlO_3$; d) $Y_{0.80}Nd_{0.20}AlO_3$ nanopowders; annealed at 1000ºC for a) - c) and 1600°C for d).**

Fig. 3 shows that the agglomeration of nanopowder grains increases with decreasing content of neodymium. For obtaining single phase composition with high Nd/(Nd+Y) molar ratio a lower temperature of annealing was required than for composition with low Nd/(Nd+Y) ratio and it was from 1000ºC to 1600ºC respectively. For this simple reason, powders annealed at higher temperatures have larger grains (Fig. 3d).

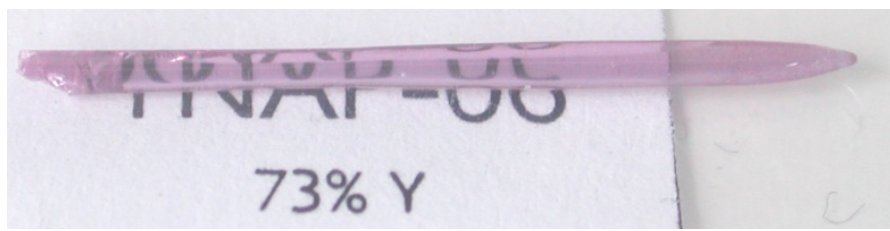

**Fig. 4: $Y_{0.73}Nd_{0.27}AlO_3$ crystal obtained by micro-pulling-down method. The length of crystal is about 25 mm.**

Homogeneous and optically clear single crystals with several centimeters length could be grown with micro-pulling-down method from powder mixtures outside the 2-phase region $0.45 \leq Nd/(Nd+Y) \leq 0.80$ that was found by X-ray measurements (Fig. 1). A crystal with $Nd/(Nd+Y) = 0.27$ is shown in Fig. 4.

DTA measurements were performed with ca. 20 compositions from the whole concentration range, including pure $YAlO_3$ and $NdAlO_3$. Pieces of the micro-pulling down crystals as well as nanopowders were used for DTA. For the powders, prior to these measurements, calcination at 1600°C in air was performed, and this treatment was found to make the samples single phase. Typically 3 heating/cooling runs were performed for every sample. Typically, only heating curves could be used for analysis, as supercooling up to ca. 50 deg was always found in the cooling runs. Some DTA curves are shown in Fig. 5. For the interpretation it should be taken into account that these curves had to be obtained from different DTA setups. This was mandatory as the Nd-rich samples could not be molten in the STA 449 that was used initially. Besides, the high number of measurements made it was necessary to replace the W-W/Re thermocouples of the sample carrier repeatedly. Unfortunately, no reliable calibration of the thermocouples is available for such high temperatures, and therefore the $T$-values of peak onsets (solidus temperatures) are expected to have experimental errors in the order of ±15 deg. Only for a few samples the liquidus could be determined as that $T$ where the DTA curve returns after passing the melting peak to the basis line (see for the $x = 0.45$ sample in Fig. 5), but the accuracy in determining this point is typically worse and an error of ±20 deg must be assumed.

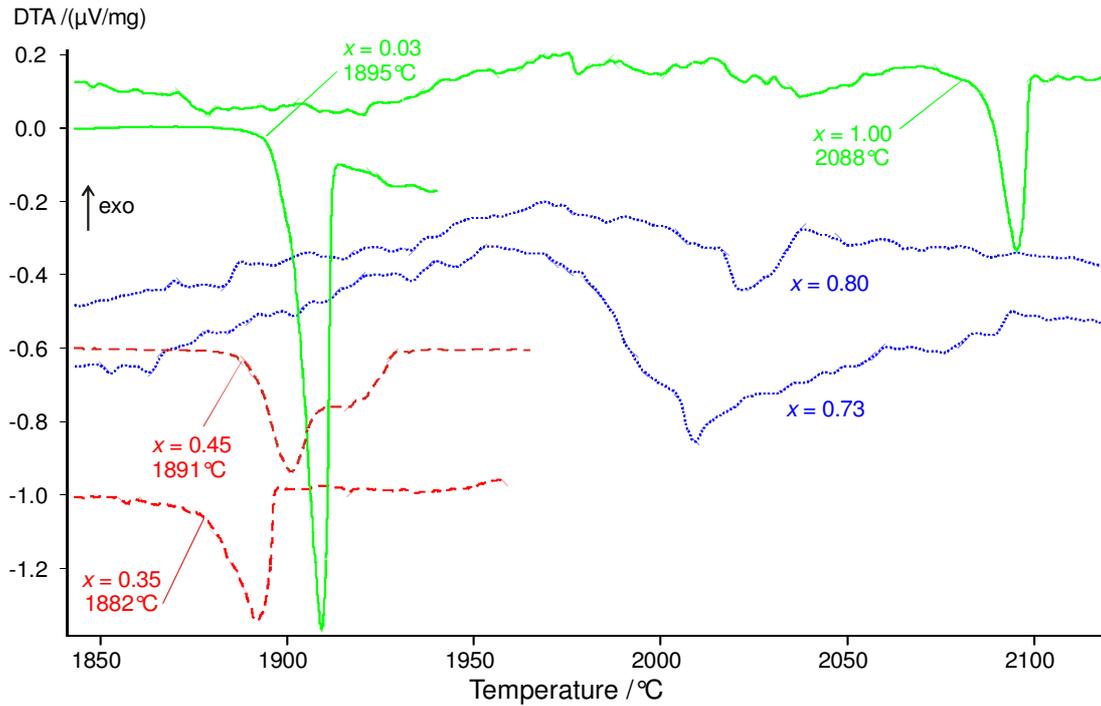

**Fig. 5: DTA heating curves for several compositions ranging from almost pure YAP (molar fraction Nd/(Nd+Y) = $x$ = 0.03) to pure NAP ($x$ = 1.00). These two compositions are represented by solid lines, compositions near the eutectic by dashed lines, and compositions in the Nd-rich solid solution range by dotted lines.**

An alternative method for the determination of the liquidus temperature in systems with supercooling from DTA measurements was proposed by Fedorov and Medvedeva [21] for eutectic systems. This method was adopted here, and was used for some compositions. It relies on "cycling" by subsequent heating/cooling runs between a constant lower temperature $T_0$ and an upper temperature $T_1$ that is stepwise increased from $T_2>T_0$ up to a limit well above the liquidus temperature:

$$T_1 = T_2 + n\Delta T$$

where $n$ = 0, 1,… is the cycle number and $\Delta T$ is the step width. This procedure is demonstrated in Fig. 6 for five subsequent cooling runs ($\Delta T$ = 5 K, $n$ = 7,…,11) of a YAlO$_3$/NdAlO$_3$ mixture with $x$ = 0.35. As long as $T_1$ is below the solidus, the whole sample remains solid and no significant thermal are observed during cooling from $T_1$ to $T_0$. If $T_1$ is just above the solidus, a small portion of the sample is molten and crystallizes during subsequent cooling to $T_0$. The crystallization occurs without significant supercooling as the major part of the sample remains solid. This is demonstrated by the bottom curve in Fig. 6 with a minor (almost invisible) exothermal peak between $T_0$+10 K and $T_0$+20 K. $T_1$ is increased for the subsequent cycles and consequently larger portions of the sample are molten. The peak area $A$ during cooling increases from cycle to cycle – as long as the upper temperature $T_1$ is within the 2-phase field between solidus and liquidus. If $T_1$ passes over the liquidus, $A$ remains approximately constant, as a constant amount of matter crystallizes. (Actually, it may fluctuate here somewhat because the heat of crystallization depends on temperature $T$, and the crystallization temperature is often not constant due to supercooling.)

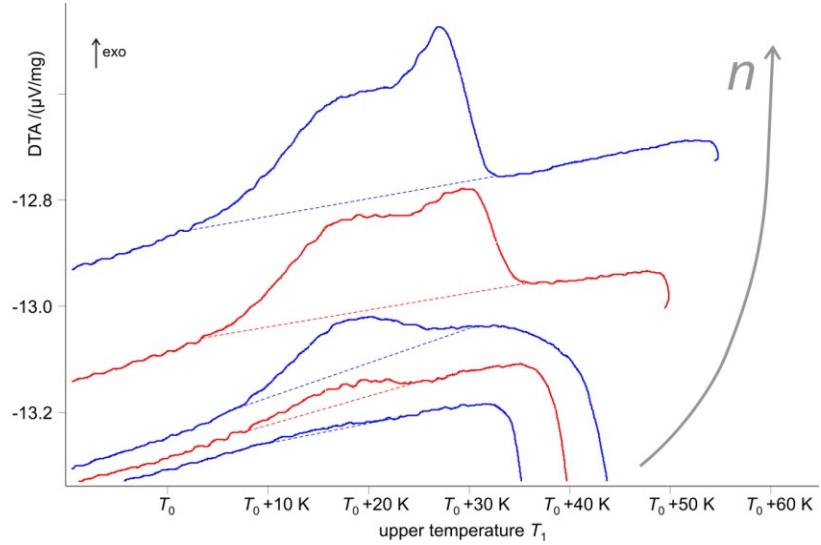

**Fig. 6: DTA cooling curves from a "cycling" measurement of $Y_{0.65}Nd_{0.35}AlO_3$ for the determination of the liquidus temperature. The subsequent cycles from $T_0+35$ K (bottom) to $T_0+55$ K (top) are shown here.**

For some compositions such "cycling" DTA measurements were used to determine the width of the 2-phase field between solidus and liquidus. Fig. 7 shows that plots $A(T_1)$ are rising from zero to a saturation. The characteristic of this function depends on the slope of the solidus and liquidus curves, as well as several on experimental parameters such as heat transport inside the DTA sample carrier. It is beyond the scope of this article to give a quantitative description, but it proved that a 3-paramter sigmoid function

$$A(T_1) = \frac{a}{1 + \exp\left[-\frac{T_1 - b}{c}\right]}$$

proved to be a satisfactory description. Here the parameter $a$ gives the saturation level, $b$ is the temperature were 50% of the saturation is reached, and $c$ describes the width of the $T_1$-range where $A(T_1)$ rises to its maximum. One can show easily that the peak area rises from 10% to 90% of its maximum value within 4.4×$c$. This $T_1$-span is marked for the two fits in Fig. 7 by horizontal bars.

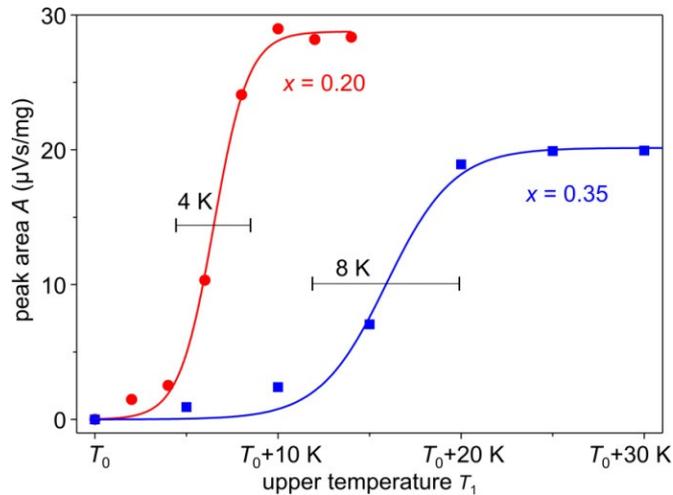

**Fig. 7: Peak areas from "cycling" two $Y_2O_3$-rich mixtures, together with 3-parameter sigmoid fits that were used for the determination of transient width (10% to 90%).**

It turned out that the melting onset temperatures drop from pure NAP ($T_f$ = 2100°C) to mixtures with 65% NAP and 35% YAP, where onsets around 1880-1890°C are observed. In this concentration range the 2-phase field between solidus and liquidus is up to 140 K broad. Not before 10% NAP/90% YAP, the onsets start to rise again slightly, up to the melting point of pure YAP ($T_f$ = 1900°C). All melting peaks from pure YAP ($x$ = 0) to $x \approx 0.35$ are narrow (< 25 K) and all peaks from $x$ = 0 to $x \approx 0.65$ have onsets in the range 1880-1900°C.

It was mentioned above that the melting peak onset of pure YAP is ca. 10 K higher than that of YAP rich mixtures (10-30% NAP). Besides Fig. 7 shows that in the middle of this region the width of the melting peaks is very small; indeed it is not broader than that of pure YAP or NAP, respectively.

One can conclude that NAP as well YAP are congruently melting compounds. A composition near $Y_{0.80}Nd_{0.20}AlO_3$ melts congruently too, as the binary phase diagram has an azeotrope point, with a common minimum of the solidus and liquidus curves. (It should be noted that a local minimum of both curves is only possible if they meet in the minimum at one point.) Solid solutions with larger NAP concentration 0.30 < $x$ < 0.65 are melting peritectically. The phase diagram is shown in Fig. 8.

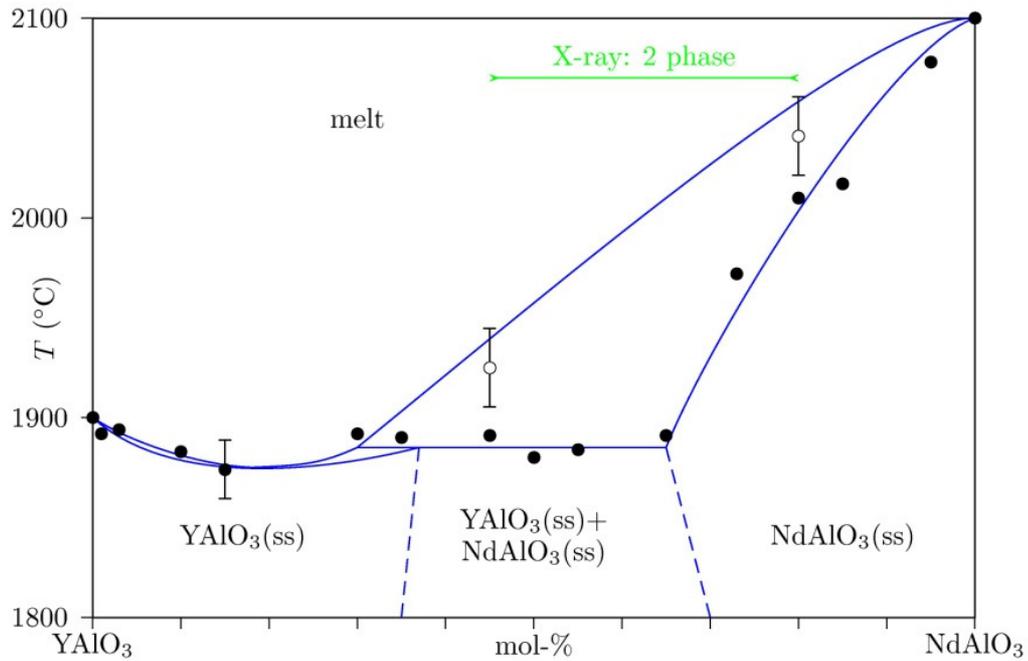

**Fig. 8: The pseudo-binary phase diagram. The boundaries of the phase field YAlO$_3$(ss)/NdAlO$_3$(ss) could only be estimated. Onsets of some melting peaks are shown by filled symbols and offsets (liquidus) by hollow symbols. X-ray diffractograms showed 2 phases in the concentration range that is indicated at the top of the figure.**

For compositions left from the green line in top of Fig. 8 (< 45% NAP) X-ray diffraction spectra (Fig. 1) showed only peaks related to the orthorhombic YAP phase, and right from this line (> 80% NAP) only the rhombohedral NAP phase was found. This limit is close to the limit concentration ≈70% NAP where the dependence of the unit cell volume from concentration becomes nonlinear (Fig. 2). The DTA results suggest a solubility limit (Nd rich end of the eutectic line) near 65% NAP, but it must be taken into account that DTA observes the equilibrium at the eutectic temperature (1900°C) whereas the X-ray measurements are performed at room temperature and represent the frozen in state at some not well defined $T < T_f$.

## 4. Conclusions

The system YAlO$_3$-NdAlO$_3$ is pseudo-binary and contains one liquid solution phase (melt) and two solid solution phases YAlO$_3$(ss) and NdAlO$_3$(ss). The YAlO$_3$(ss) phase has an azeotropic point near 20% NdAlO$_3$ doping with a melting point that is ca. 20 degrees below that of pure YAlO$_3$. For all YAlO$_3$-rich solid solutions the 2-phase region between solidus and liquidus is observed, and consequently bulk crystal growth from the melt is possible there without significant segregation, such as the crystal with $x = 0.27$ that is shown in Fig. 4. This conclusion agrees with the observation of Bagdasarov (1969) that high Nd$^{3+}$ concentrations together with better homogeneity can be obtained with YAlO$_3$, as the distribution coefficient of this doping ion is close to unity[12]. The behavior of the perovskite phase is more beneficial than that of the garnet phase $Y_3Al_5O_{12}$ where the solubility of $Nd^{3+}$ is also high, but segregation is so strong that doping levels exceeding a few percent are not accessible for melt growth[22].